# Physical Mechanism of Superconductivity Part II
## Superconductivity and Superfluidity

**Yu-Ru Ge, Xin Zhao, Hong Zhao and Xue-Shu Zhao**

**Abstract**.  The transition mechanism of metal - insulator in metal oxides is discussed in detail, which is a part of the mechanism of superconductivity. Through the study of magic-angle twisted bilayer graphene superconductor and other new findings on superconductivity, we further demonstrate that the physical mechanism of superconductivity proposed in Part 1 is the only correct way to handle the properties of superconductivity in various materials. We propose that superfluid helium consists of normal liquid helium mixed with high-energy helium atoms. Based on this new model, all peculiar features discovered in superfluid helium can be truly understood, such as the climb of superfluid helium on the container's wall, the fountain effect, the kapitza conductance, the discontinuity of specific heat capacity at phase transition, and the maintaining mass current in a ring-shaped container. We demonstrate that high-energy particles play a driving force role in both superconductors and superfluid helium, and therefore dominate their properties.

## I  Introduction

Physical laws provide us a consistent and coherent picture of the matter world. As correct theories of superconductivity or superfluidity, they should be rigorously in unison with the physical laws and be able to explain the properties of all superconducting materials and superfluid helium consistently, respectively.   Even if there is only one piece of hard evidence against its statement, the credibility of the theory should be seriously destroyed. The physical mechanism of superconductivity proposed in Part I [1] was published ten years ago, and all physical properties discovered previously have been elucidated successfully.

In this article, we will first discuss some physical concept such as metal-insulator transition in metal oxides and about Mott insulator. We will then discuss some new findings in superconductivity, including strange phenomena observed in superconducting current and voltage, the effect of strained films on superconductivity, light-induced superconductivity, and magic angle-twisted bilayer graphene superconductor. Based on the discussions of the physical properties observed in superconductors so far, we further demonstrate that the physical mechanism of superconductivity proposed in part I [1] is the only correct way to understand the properties of superconductors in all materials.

Superfluid helium and superconductivity do have some features in common, which means that they should have similar physical mechanisms. For instance, they both have an order-disorder transition in which heat capacity undergoes discontinuous changes. In addition, the ability of superfluid helium to maintain circulating mass current in a ring-shaped container is closely similar to the persistent electric current in a superconducting ring. Based on the physical mechanism of superconductivity in part I [1], we propose that the coherent mass current in both superconductor and superfluid helium is generated by dynamic process caused by high-energy particles.

We propose that superfluid helium consists of normal liquid helium mixed with high-energy helium atoms, which are generated during a phase transition. Keeping in mind with this new model, all the peculiar features discovered in superfluid, such as the climbing of superfluid helium on container walls, the fountain effect, and the maintaining circulating mass current in a ring-shaped container, become quite reasonable. They can all be completely understood in terms of classical physics. It should be emphasized that high-energy particles in both superconductors and superfluid helium act as the driving force and dominate their properties.

## II  On Superconductivity

For convenience, we need to review some key points of the physical mechanism of superconductivity in Part I [1]. A superconducting material must satisfy the following three necessary conditions: First, the material must contain high- energy nonbonding electrons with a certain density requested by the coherence length, to provide energy for the generation of superconducting state. High- energy particles play the driving force role, which lies at the heart of dynamic processes such as superconductivity and superfluidity. Second, the superconducting material should possess three-dimensional potential wells lying in energy at above the Fermi level of the material. The dynamic bound states of superconducting electrons in potential wells in a given superconducting chain should have unified binding energy and symmetry. According to the type of potential well, superconductors as a whole can be divided into two groups. One of these is regarded as a conventional superconductor, in which potential wells are formed by the material's microstructures, such as crystal grains. The other group is referred to as high-$T_c$ superconductors or unconventional superconductors, in which potential wells are constructed only by the lattice structure of the corresponding materials, such as $CuO_6$ octahedrons and $CuO_5$ pyramids in cuprates. Finally, for the material to be metal in its normal state, the lowest conduction band of the superconducting material should be widely dispersed, so that it can extend beyond the height of the potential well. The symmetry of the lowest conduction band dominates the type of superconducting state. However, if the lowest conduction band of the material cannot run over the height of the potential well, then the material should exhibit the characters of an insulator in its normal state. In the history of solid-state physics, this kind of material is considered a Mott insulator [2].

On the other hand, the concept of Mott insulator originated from the challenges of transition metal oxides. Based on conventional band theory, a number of transition metal oxides should exhibit the characteristics of conductors, but in fact, they behave as experimentally determined insulators. Mott proposed in 1937 that this discrepancy could be explained by considering the mutual repulsion of free electrons, which are not taken into account in conventional band theories. According to Mott's predictions, mutual repulsion of conduction electrons could also lead to metal-insulator transition in metal oxides. As far as our knowledge goes, in entire history of condensed matter physics there has not been any proven material in which the metal-insulator (or vice versa) transition is caused merely by mutual repulsion of conducting electrons. A large number of experimental results have proven that the interactions of electrons with the atomic lattice play a dominant role in the phase transition of all condensed matter.

## 2.1 On the Metal-Insulator Transition in Metal Oxides

We proposed in Part I [1] that metal-insulator transition in metal oxides results from the fact that the maximum of the lowest conduction band of the material is confined in potential wells located at above the Fermi energy level.

In order to illustrate this mechanism more clearly, we choose following two sorts of metal oxides: one group is the high-$T_c$ superconducting metal-oxides $La_2CuO_4$ and $BaBiO_3$, and the other is the conventional superconductor $Nd_2CuO_4$. According to band theory, they should all be metallic. However, experimental results showed that the two metal oxides in the first group behave as insulators, while the compound $Nd_2CuO_4$ exhibited metallic characteristics. In fact, the difference between the two groups lies in their lattice structures. In the first group, materials $La_2CuO_4$ and $BaBiO_3$ have the octahedral lattice structure $CuO_6$ and $BiO_6$ (see Fig.1, right and center), which determine the band structures of both materials nearby the Fermi level, respectively [3].

The compound $Nd_2CuO_4$ has a lattice structure similar to that of $La_2CuO_4$, but the two $O_Z$ atoms in the unit cell of $Nd_2CuO_4$ are displaced from their apex positions onto the sits on the faces of the tetragonal cell. So that, the Cu ions in $Nd_2CuO_4$ are square planes coordinated with oxygen atoms (see Fig. 1, left), and there is no three – dimensional potential well surrounding Cu ions [3]. Based on the tight binding model, the lowest conduction band of pure $BaBiO_3$ compound is mainly composed of Bi6s – O2p$\sigma$ anti-bonding band, primarily of Bi6s extended orbital [4]. While the lowest conduction band in pure $La_2CuO_4$ and $Nd_2CuO_4$ compounds all result from Cud$x^2$-$y^2$ - O2p$\sigma$ antibonding band, which is mainly formed by the extended Cud$x^2$-$y^2$ orbital or pd$\sigma$ band. The anti-bonding states for $La_2CuO_4$ and $BaBiO_3$ compounds are confined in $CuO_6$ and $BiO_6$ octahedral potential wells, respectively [5].

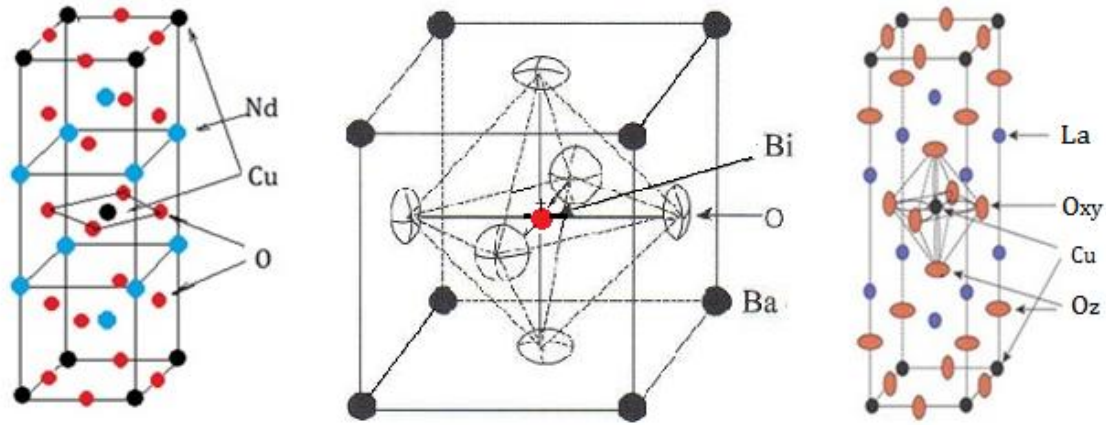

Fig. 1. The lattice structures of La2CuO4, BaBiO3 and Nd2CuO4 compounds. Based on the calculations of band theories, the three compounds should be metallic, but La2CuO4 and BaBiO3 are identified as insulators. The conflict between theoretical predictions and experimental results is caused by their lowest energy bands of La2CuO4 and BaBiO3 are confined in the $CuO_6$ and $BiO_6$ potential wells in their corresponding compounds, which cannot be taken into account in all the energy band theories we have now.

The lattice of pure $BaBiO_3$ compound has a cubic symmetry, meaning that charge density on the six ellipsoids of $BiO_6$ potential wells must be the same (Fig. 1, center). Therefore, it follows that the potential height in the cubic face directions should all have the same value, which is about 2eV measured by optical experiments. While the highest energy states of the Bi6s-O2p$\sigma$ antibonding band is only 1.7eV above the Fermi level, thus it is confined strongly in $BiO_6$ potential well [1]. Thereby, there is no doubt that the normal state of pure $BaBiO_3$ shows insulator properties.

On the other hand, the pure $La_2CuO_4$ compound has the axial symmetry. The potential height in Oz direction is greater than 6eV due to the 2$^-$ valence state in Oz ions. However, through optical conductivity measurement, the potential height in $CuO_2$ planes is about 2eV. Thus, the $La_2CuO_4$ compound must show strong two-dimensional characteristic. In addition, based on the band structure calculated by tight binding model, the maximum of pd$\sigma$ antibonding band in the pure $La_2CuO_4$ is about 1.8eV above the Fermi level (for detail please read Part I [1]). Therefore, it is not surprising that the pure $La_2CuO_4$ compound behaves as insulator. However, there are no three-dimensional potential wells surrounding Cu ions in $Nd_2CuO_4$ compound (shown in Figure 3, left). No matter how high the Cud$x^2$-$y^2$ – Opd$\sigma$ antibonding band goes up, $Nd_2CuO_4$ compound always characters as a metal.

Therefore, we conclude that the contradiction between theoretical predictions and experimental results for the characteristics of the pure $La_2CuO_4$ and $BaBiO_3$ compounds is due to the existence of $CuO_6$ and $BiO_6$ potential wells in the corresponding compounds, which cannot be taken into account in all current energy band theories. The common rule can be drawn is that if the height of Cu(Bi)$O_6$ potential wells is higher than the maximum value of the lowest conduction band in the corresponding material, then the compound will show insulating characteristics in its normal state. Another factor that all band theories fail to account for is the effect of the high-energy nonbonding electrons on the physical properties of superconducting materials. Two effects that band theories cannot take into consideration are located at the heart of superconductivity.

The hole doping has two functions: one is to reduce the free energy density of the doped unit cell. For instance, when a trivalent La atom in $La_2CuO_4$ is replaced by a divalent atom Ba, the compound will lose a high-energy La5d electron in the doped unit cell. As a result, the free energy density in the doped cell becomes lower than that of its surrounding undoped cells. When the pressure and temperature are held constant, the Helmholtz free energy can be written as $dF = - PdV$ or $dF/dV = - P$. It follows that the decrease of free energy density in doped unit cell must undergo a compressive pressure (-P) caused by its surrounding crystal field, which consequently leads to a decrease of the Cu – Oxy bond length in the doped cell. Following bond length shrinking, the energy of the antibonding state Cud$x^2$-$y^2$ in the doped cell will move upwards, and runs over the height of potential well in $CuO_2$ planes. If the average distance between two nearest doped cells is smaller than the free path of electrons in $CuO_2$ planes. Then, the hole doped material $La_{2-x}Ba_xCuO_4$ changes from insulator to a metal in $CuO_2$ planes.

Another function of hole-doping is to adjust charge densities of superconducting electrons to a specific value required by coherence lengths. The coherence lengths in cuprate compounds must satisfy the relation $\xi = 2na$ in $CuO_2$ planes, here n = 1, 2… integer, a is lattice constant. Supposing that the coherence length equals 4a (about 1.6nm) in $La_{2-x}Ba_xCuO_4$ compound, and then the corresponding optimal concentration of divalent Ba atom should be x = 0.125. That is, there is one superconducting electron in every four-unit cells in a given superconducting chain.

It follows from the discussions above that the insulator-metal transition in both compounds $La_2CuO_4$ and $BaBiO_3$ has nothing to do with mutual repulsions of conducting electrons. After the Ba doped $La_2CuO_4$ compound ($La_{2-x}Ba_xCuO_4$) completes the insulator- metal transition, the $5_d$ electron on another La atom in the doped cell has an opportunity to make the transition from its atomic orbit to the $d_{x2-y2}$ conduction band laying at the top of $CuO_6$ potential wells, becoming conducting electron with a kinetic energy equal to its transition energy. Therefore, the current of superconductor is conducted by the electrons with high kinetic energy at the conduction band far above from Fermi level.

It is worth noting here that according to the calculations of band theories, the Fermi level in $La_2CuO_4$ compound is located at the energy levels of Cu- $O_{xy}$ and Cu- $O_z$ bond states. Therefore, the electrons of the Fermi level are the electrons that combine with the oxygen ions of the potential well $CuO_6$ [5]. Therefore, the potential well formed by $CuO_6$ must be located at energies above the Fermi level. This conclusion should be true for all of superconducting materials.

As temperature is lowered below $T_c$, the high – energy electrons in the conduction band have chance to trap themselves into the doped unit cell, where the free energy density is lower compared with the undoped unit cell. According to carrier- induced dynamic effect (CIDSE) in Part 1 [1], the high-energy $La_{5d}$ electron in the $d_{x_2-y_2}$ conduction band will use its kinetic energy to push the four nearest $O_{xy}$ ions in $CuO_2$ plane outward, leading an increase in Cu- $O_{xy}$ bond length, which in turn brings the antibonding band $d_{x_2-y_2}$ back into the $CuO_6$ potential wells. Then the conducting electron becomes dynamic bound state in potential wells, and the binding energy of the trapped electrons dominates the superconducting transition temperature in the corresponding material.

The concept of dynamic bound state is completely different from the conventional bound state defined in solid-state physics. The dynamic bound state, as defined here, is formed by high energy conducting electron, using its kinetic energy to push its surrounding lattice moving outward, and then trapping itself in a potential well. Thus, the dynamic bound states must be located in conduction bands. However, the conventional bound state is caused by the electrostatic interaction between electrons and their surrounding lattices. Usually the conventional bound states are located in valence bands or band gap.

When all of high-energy conducting electrons become dynamic bound states in their corresponding potential wells, the material then enters into its superconducting state. Once superconducting state is achieved, the high-energy electron trapped in potential well $CuO_6$ will push the nearest four oxygen ions in $CuO_2$ plane moving outward, and at the same time, bring two $O_z$ ions moving inward. Consequently, the entire compound undergoes an order-disorder transition. In company with this transition, the specific heat of the material also experiences discontinuous changes at the transition point. It is clearly that superconducting compounds undergo two kinds of transitions in the process of achieving superconductors: one is a metal-insulator transition and the other is an order-disorder transition.

However, the mainstream of researches on superconductivity has attributed the metal – insulator transition to Mott insulator caused by the mutual repulsion of conducting electrons. Obviously, this prediction is unacceptable, for instance, there is only one conducting electron for every four unit cells in $La_{2-x}Ba_xCuO_4$ compound for x = 0.125. A large number of the charged oxygen ions between them must strongly shield the mutual repulsion between the two conducting electrons. So that, in any case the mutual repulsion among conducting electrons cannot play the driving force role for the metal- insulator transition during the superconducting processes.

Thus, we can draw the conclusion that the onset of superconducting state in any material must be accompanied by lattice distortions, which in turn will transfer the lattice structure of superconducting material from order to disorder, or some other lower symmetry structure, depending on the symmetry of dynamic bound states. The order-disorder transition induced by dynamic bound state of high- energy electrons does not permanently change crystalline phase of the superconducting materials. During the superconducting processes, the superconducting material will transfer periodically between its crystalline phase and disorder phase. In other words, no matter how strong the dynamic distortion is, both normal state and superconducting state (or insulating state) must exist in the same crystalline phase. In superconducting state, lattice distortion wave or charge density wave is generated periodically along superconducting chain; the distortion wave has the same wavelength as the superconducting electrons. Under an action of an electric field, superconducting electrons move coherently with lattice distortion wave or charge density wave, and periodically exchange their excitation energy with chain lattice. That is, superconducting electrons move alternately between their dynamic bound state and conducting state. So that, superconducting electrons cannot be scattered by chain lattice, and super-current persists over time. Therefore, the intrinsic feature of superconductivity is to generate an oscillating current under DC voltage. The wavelength of oscillating current or the charge density wavelength equals the coherence length of superconducting electrons, which must equal an even- number times lattice constant in cuprates.

Interestingly, K. Onner in Leiden laboratory discovered superconductivity a century ago, now Milan Allan's group in the same laboratory has discovered that charges trapped in cuprate materials conduct current with zero resistance, which is a key nature of superconductivity [6].

In 2017, based on the experiments of transient photon conductivity on the under-doped $BaPb_{1-x}Bi_xO_3$, scientists from Max Planck institute provided the reliable evidence that superconductivity and charge- density waves coexist in high $T_c$ - superconductors [7]. It follows that superconducting electrons must move coherently with charge density wave.

**2.2 Strange Phenomena observed in superconducting current and voltages**

A couple of research groups have found that in cuprate superconductors, the conducting electrons are not enough to carry the measured current, that is, the measured charge current is always greater than the current conducted by superconducting electrons [8]. This is means that something other than electrons also can conduct current.

We have known that superconductors generate alternating current under DC voltage. The lattice distortion wave or charge density wave caused by charged oxygen ions move coherently with superconducting electrons. Charge density wave induces periodic change of the electric displacement field D in superconductors. Changes of the electric displacement field over time will induce a current. Maxwell named it "displacement current", and its density is equal to $\frac{\partial}{\partial t}\Phi$, where $\Phi$ is electric flux, $\Phi = \iint D\, dx$, the surface integral of D. That is to say, in any material that conducts alternating current, the total current measured on external circuit should be the sum of the conducting current carried by the moving electrons and the displacement current induced by the time-varying displacement electric field. The faster the change, the greater the intensity is. This discovery further proves that the intrinsic nature of superconductivity is to generate oscillating current under dc voltage.

Scientists at Brookhaven National Laboratory observed that when a superconducting current flows through a cuprate superconductor, an unexpected spontaneous voltage appears and is perpendicular to the current [9]. They studied how this spontaneous voltage depends on current direction, chemical composition and crystal structure. Since they found that the phenomenon showed up repeatedly, they concluded beyond a doubt that the effect is intrinsic in the high-$T_c$ cuprate itself, and its origin is purely electronic.

We pointed out in Part I [1], that in a superconducting chain of cuprate, the potential well between two expanded potential wells caused by trapped electrons is subject to compressive strain. Under compressive strain aligned with the chain direction, the $d_{x^2-y^2}$ anti-bonding band will split into two bands; one of them is the energy band formed by the Cu-Oxy bonds perpendicular to the chain direction. As its bond length increases, its energy will move down below the height of the potential well. Thereby, electrons cannot move along the direction perpendicular to the superconducting chain. Another energy band formed by Cu-Oxy bonds parallel to the superconducting chain direction will move its energy upwards beyond the height of the potential well caused by the decrease in bond length. Thus, the superconducting current can only be conducted along superconducting chains. As the supercurrent is removed, the superconductor will return to its normal state. During this process, the lower potential energy perpendicular to the current direction (or the superconducting chains) must rise up toward its normal state. At the same time, a large number of the charged oxygen ions will move in the direction perpendicular to the supercurrent and return to their equilibrium position, which will also induce a voltage in the direction perpendicular to supercurrent. Therefore, it is not surprising to observe a spontaneous voltage appearing perpendicular to the current when supercurrent flows out of cuprate.

## 2.3 Strain Films of Superconducting Materials

The research group at University of Wisconsin - Madison created the materials by growing thin crystal film on two different support substrates. One is a compressed thin film of $La_{1.85}Sr_{0.15}CuO_4$, and the other is a tensile-strained film. Experimental results for both strained films show that tensile-strained films require much lower temperature (7K) than compressed strain films (31K) to become superconductors. The critical temperature of both strained films is well below 40 K, which is the critical temperature of unstrained bulk materials [10].

We pointed out in Part 1 that, if a compressive or tensile type strain is introduced into a given superconductor, the superconducting process in the material is unavoidably destroyed, because the introduced strain may seriously interrupt the coherent movement of superconducting electrons with lattice deformation waves. Additionally, the introduced strain will change the symmetry and volume of the potential wells in the strained region. It follows that the binding energy of superconducting electrons in the strained region should be non-uniform from one place to another, so the unified superconducting state cannot be maintained in the strained materials.

However, if the strain is uniformly distributed throughout the material, and all potential wells have the same volume and symmetry, superconductivity can be achieved just as the films made in this experiment. They also found that the substantial decrease of superconducting $T_c$ in the tensile-strained films is caused by the serious loss of oxygen atoms in the tensile-strained films, so it is reasonable to consider that there are no unified CuO6 potential wells in the tensile strained films due to the loss of oxygen atoms. In fact, measured $T_c$ (7K) in the tensile strained film is due to other microstructures rather than the $CuO_6$ potential wells. That is, tensile-strained films no longer belong to the realm of high-$T_c$ superconductors, but become conventional superconductors.

The experimental results prove that there do have the unified $CuO_6$ potential wells in compressed-strain films. Thus, the decrease of superconducting $T_c$ in the compressed-strain films is undoubtedly caused by the compressed strain. Since compressed strain makes the Cu-O bonds in $CuO_6$ potential wells stronger than the Cu – O bonds in the ordinary $La_{1.85}Sr_{0.15}CuO_4$ compound. As a common rule, strain, no matter tensile or compressive, can only change the energy of Cu-O bonds, but not the energy of the nonbonding La 5d electrons. When the high-energy La5d electrons trap themselves into $CuO_6$ potential wells by pushing surrounding oxygen ions outwards, during the process, the high-energy electrons must withstand the resistances caused by two parts, one is from the normal Cu-O bonds and the other is from the compressed strain. As a result, the binding energy induced by superconducting electrons in the compressive strain potential well must be less than that they can obtain in normal materials. Therefore, it is reasonable that the transition temperature in the compressed strained film is lower than that observed in the same normal material.

## 2.4  light-Induced Superconductivity

In recent years, a number of light-induced superconductivity experiments have revealed that shining intense pulses of infrared light at superconductors can make them work at a much higher temperature than they can do normally. These findings seem to open a new way to make superconductors work at higher temperature, even at room temperature. However, the experimental results on high-$T_c$ superconductors, including cuprates, iron-base and $K_3C_{60}$ superconductors, show that the light- induced superconducting state lasts only a couple of picoseconds [11].

In addition to extremely short lifetime, all light- induced superconducting states are accompanied by remarkable lattice modulation. On the other hand, the light excited electrons cannot directly trigger the superconducting state; it can only be achieved after the excited hot carriers return to ground state [12].

The mechanism of light-induced superconductivity is still under debate. One point of view believes that the nonlinear excitation of certain phonons caused by intense infrared pulses is directly related to the light-induced superconducting state. Another view is that the large lattice modulation accompanied by the induced superconducting state is caused by the change in the distribution of the electronic system [13].

Based on the physical mechanism of superconductivity [1], light-induced superconductivity at high temperature is naturally expected. All photon excitations from infrared to ultraviolet can instantaneously drive shallow bound electrons or free electrons to high-energy non-equilibrium states. For instance, if a dynamically bound electron in a potential well of a cuprate superconductor is excited to a transient state by a photon with energy equal to 1 eV, then the electron will obtain 1 eV of energy from the photon. When the excited electron returns to the potential well, it will use the gained 1eV energy to push its surrounding lattice outwards further than before. In this case, the electron will generate a transient dynamic bound state with a binding energy larger than before. It follows that the transition temperature of the superconductor increases instantaneously. Since the additional binding energy obtained by the bound electron is generated by the external excitation rather than inherent energy, the additional lattice distortion induced by the excitation energy must be converted into harmonic vibrations (phonons) to bring out the excitation energy. Here, we can imagine that the so-called "nonlinear phonons" may mean the lattice distortion caused by the excited electrons. Thus, we can see that no matter which method is used, light-induced superconductivity is destined to have a very short lifetime. Perhaps the lifetime can only persist in a couple of oscillation periods.

When superconducting electrons are excited by light from their dynamic bound state to their transient excited state, the superconducting process cannot begin immediately. Because the excited- electrons do not have a unified dynamic bound state associated with them. They must return to the potential wells to generate unified dynamic bound state before the superconducting process can start coherently.

On the contrary, for any superconductor, when its temperature is below $T_c$, a superconducting state is formed, in which superconducting electrons become dynamic bound state in their corresponding potential wells. In this case, if some atoms on the potential wells form "nonlinear phonons" in the process of intense infrared light exit, then the superconducting process must be destroyed, because the unified dynamic bound state that should has same energy and symmetry as the original superconducting state is broken.

The effect of light-induced superconductivity directly proves that there is no Cooper pair in superconductors. The binding energy of Cooper pairs in room temperature superconductors is only about 25 meV. However, the photon energies used in the photo excitation experiments ranged from 155 meV to 1.6 eV, much higher than the binding energy of Cooper pairs in the corresponding material. In addition, by exciting a superconductor with a photon energy that exceeds the Cooper pair binding energy, the paired electrons should be split apart and superconductivity should be destroyed. But in fact, under the excitation of light, not only the superconductivity is not destroyed, but also the superconducting transition temperature increases instead. Therefore, we conclude that Cooper pair is not the real mechanism that causes superconductivity; In fact, there are no such things in superconductors.

## 2.5  Magic-Angle Twisted- Bilayer- Graphene (MATBG)

In 2011, Rafi Bistritzer and Allan H. MacDonald first calculated the electronic structure of a twisted two -layer graphene system [14] and found that the coupling between two layers of graphene is stronger. For a discrete set of magic angles, the Dirac velocity vanishes and the lowest moiré band flattens.

The discovery of the flattening of the lowest conduction band of magic-angle twisted-bilayer- graphene (MATBG) laid the foundation for the afterwards finding superconductivity in MATBG. Seven years later, in 2018, an MIT research team reported a method to achieve intrinsic unconventional superconductivity in 2D superlattices by stacking two graphene sheets with a magic twist angle of $1.1^o$ [15]. They also found that twisted- bilayer- graphene exhibits ultra-flat band near charge neutrality, which lead to a correlated insulating state when half-filled. After electrostatic doping away from these correlated insulating states, they observed a tunable zero-resistance state with a critical temperature of 1.7 K. Temperature-density phase diagrams shows similarity to cuprates. They attributed MTABG to the strongest coupling superconductors, based on a small Fermi surface near correlated insulating phase and a relative high $T_c$,

This surprising discovery has prompted people to conduct in-depth researches on its underlying physical mechanism. The twisted- bilayer- graphene systems containing only carbon atoms can potentially provide a simpler and highly tunable platform for studying the root cause of high-$T_c$ superconductivity in cuprates. The MATBG superlattice creates a Moiré lattice with period L,

$$L = a / (2\sin\frac{\theta}{2}) \qquad (1)$$

Where a = 0.246 nm is the graphene lattice constant, θ is the twist angle. At magic angle $\theta = 1.07^0$, get L = 13.2 nm. The unit cell of a Moiré lattice consists of three stacking regions.

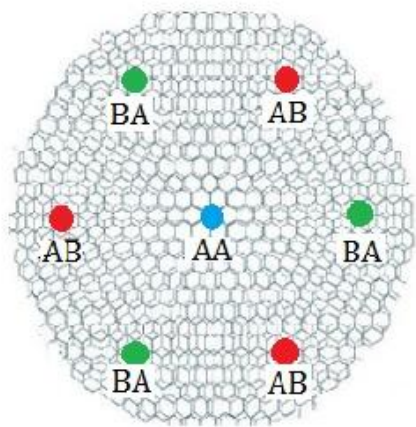

Fig. 2. The unit cell of Moire superlattice is consisted of three stacking regions. The AA stacking region is located at the center of the Moire unit cell. The other two Bernal stacking AB and BA occupy the remained region of the unit cell.

To achieve superconductivity, the most important is the AA stacking region located at the center of the Moiré unit cell (Fig. 2), where each top-layer atom is located directly above the corresponding atom in the lower layer; the AA stacking region occupies very small area, about few unit cells of a single graphene layer. Even with zero doping, the AA stacking region is a metal, but quite unstable. The other stacking regions AB and BA (Bernal stacking) occupy the remaining area of the Moiré unit cell. In the AB (BA) stacking region, each top-layer atom in the A(B) sub-lattice is directly above the B(A) atom in the lower layer, while top-layer B(A) atom has no partner in the lower-layer, as shown in Figure 2. The Bernal AB (BA) stacking lattice is robust and stable compared to the AA stacking region.

According to the carrier-induced dynamic strain effect (CIDSE) theory given in Part I [1], the key factor for realizing superconductivity is the high-energy nonbonding electrons, which provide energy for forming superconducting state. For MATBG, the high-energy electrons should be the electrons injected through the bias voltage of STM experiment. If an electron is injected into a bulk crystal, it will continuously interact with crystal lattice, and transfer all of its energy to phonons (thermal energy), then relax to the Fermi level of the corresponding material. However, the electrons injected into MATBG superlattice have no chance to release their energy to the bi-layer lattice. The injected electrons will directly trap themselves into the AA stacking regions, where they become the dynamic bound state. The second necessary condition for realizing superconductivity is that superconducting materials must possess three- dimensional potential wells located at above Fermi level of the corresponding material. The function of potential wells is to confine the high-energy electron into a small area in which the high-energy electron becomes a dynamic bound state through dynamic interaction with its surrounding lattice. The binding energy of the trapped electron dominates the critical temperature of the superconducting material.

The tiny AA stacking region and its flat conduction band together constitute perfect potential well in MATBG super-lattice. The flat lowest conduction band means that the Fermi velocity of electrons in AA stacking regions should be zero, in other words, the injected electrons are confined in AA stacking regions. A number of experimental results show that the peak of local density of electron state is located in AA stacking regions, rather than AB or BA regions [16]. So that, all of injected- electrons will directly fall into AA stacking regions. On the other hand, the tight binding calculations show that the local density of electron state close to charge neutrality are also located in AA stacking regions, and the flat conduction band in the AA stacking regions is mainly composed of the anti-bonding states of pp$\sigma$ bonds of graphene sheet [15, 17].

What we need to emphasize here is that, by definition, the high-energy non-bonding electrons are considered to be electrons inherent in superconducting materials rather than the electrons injected from external circuit. In principle, the injected electrons cannot establish a stable dynamic bound state in a robust potential wells, because they must interact with lattice to release all energy they gained from external circuit to phonons (thermal energy). However, the tight-binding band structure shows that for zero doping, the Fermi surfaces of the two flat bands unexpectedly coincide. So that, arbitrarily small electronic interaction can disrupt the lattice of AA stacking regions, destabilize such degenerate spectrum, and generating an energy gap [17].

Assuming that an electron is injected into the MATBG super-lattice through a bias voltage of 200meV, just as the case used in STM experiment [16]. The injected electron will directly fall into an AA attacking potential well, where the Fermi velocity or translation speed of the injected electron is equal to zero. That is, the kinetic energy of the electron is vanished.

Therefore, the injected electron stores all its energy 200 meV in its wave function as free energy. In addition, the wave function of the flat conduction band in the AA stacking region is mainly consisted of the antibonding states of pp$\sigma$. Due to the highly metastable of AA stacking lattice, the energy stored in the p type wave function of the injected electrons will create tensile strains aligned with the energetically preferable direction, which is the crystallographic axis of the Moiré superlattice. Thus, the injected electrons have no chance to release their energy to phonons. This is why we regard the AA stacking regions as perfectly unparalleled potential wells for the MAYBG superlattice.

It is worth noting that the trapped electron in AA stacking region must occupy the space between the two graphene layers because where the energy is the lowest compared to the two surface layers. According to the carrier-induced strain model [1], the injected electron would use one- third of its energy to generate lattice strain and use two- third of its energy to maintain the strain. The tensile strain caused by the trapped electron in AA stacking regions would elongate the length of carbon bonds in AA stacking region in the direction along with crystallographic axis of the Moiré superlattice. Elongated carbon bonds should reduce the conduction band energy to below Fermi level along the crystallographic axis of Moiré superlattice. Consequently, the injected electrons will be trapped into their dynamic bound states in AA stacking regions. The dynamic bound states lying below Fermi level are just the origin of the lower band observed by Yuhang Jiang et al [16]. In addition, the tensile strain aligned with a crystallographic axis must be accompanied by a compressed strain perpendicular to the axis, which will lead to shrink of carbon bonds in this direction. As a result, the energy levels perpendicular to crystallographic axis are forced to move up, which happened to be the upper hole's band measured by Jiang's group [16].

According to the criteria for achieving superconductivity, the coherence lengths in MATBG superlattice should be equal to the distance, which must be an even multiple of the distance between two nearest adjacent centers of AA stacking regions along crystallographic axis. Assume that the period of a given moiré unit cell equals L = 13.2nm for twist angle $\theta$ = $1.07^0$, then the distance between two nearest adjacent centers of AA stacking regions should be 22.86 nm, which is also equal to the size of moiré unit cell. Thus, the shortest coherence length in MAYBG superlattice should be 45.7nm (Figure 3), which corresponds to a density of injected electron $1.15 \times 10^{11}$ cm$^{-2}$. The next shorter coherence length is 90.14nm, corresponding to a density of injected electron equal to $0.5 \times 10^{11}$ cm$^{-2}$.

Suppose that we have a superconducting state with the shortest coherence length 45.7nm in the MATBG superlattice, and then there is a dynamic bound state in every other moiré unit cell along a crystallographic axis in the MATBG superlattice. Perhaps this is the most possible case for p type superconductors. Because the uniaxial strain only slightly reduce the volume of the material, so the energy level shift under uniaxial strain is usually very small. The tensile strain caused by electron trapping into an AA stacking region will produce the compressed strain to the two nearest neighbors in superconducting chains. As a result, the AA stacking region between the two tensile strained cells (the two dynamic bound states) undergoes compressive strain, which shifts the energy level aligned with the chain direction upwards to conducting states. While the energy level in the direction perpendicular to the chain direction moves down below the Fermi level, and behaves as an insulator. Now we have a completely superconducting state, in which all electrons stay in their dynamic bound states, and then the whole superconductor becomes an insulator. This is the only way for all of superconductors; both conventional and unconventional must follow. That is, the superconductivity must start from an insulating state. Under dc voltage, the bound electrons in AA stacking regions will obtain excitation energy from electric field and become conduction electrons. The conducting electrons carrying all their energies with them move coherently with the lattice distortion wave to their nearby empty AA stacking regions, where all conducting electrons become dynamic bound electrons again. Then electrons will continue to repeat the trapping and moving processes, creating a superconducting current along the superconducting chain.

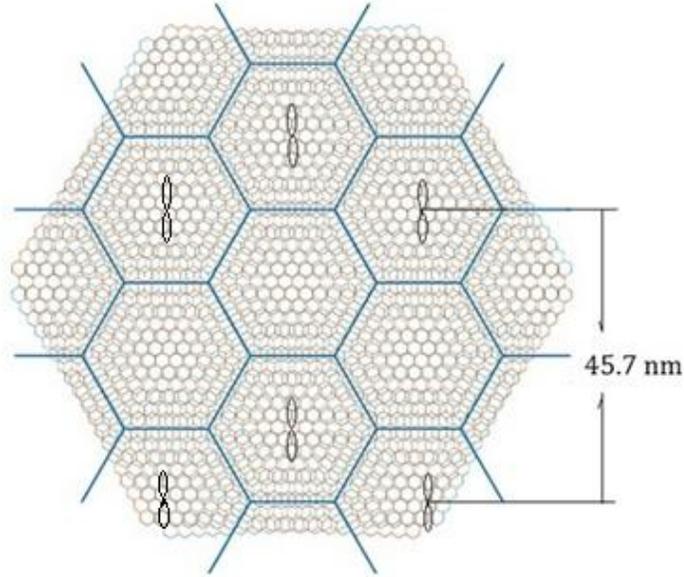

Fig.3. The injected electrons are bound into the AA stacking regions due to the flat condution band in AA stacking regions and the instability of AA stacking lattice. The flat conduction band is consisted of pp$\sigma$ antibonding obitals. The electron charge distribution for an electron bound in the extended p obital would be aligned with the symmetry axis of the moire superlattice. The distance between two nearest bound states equals the coherence length of 45.7 nm.

It is worth noticing that, in high-$T_c$ cuprates, the high-energy electrons forming the dynamic bound states exist inherently in the cuprates, thus the critical transition temperature cannot be affected by external conditions. However, in MATBG superlattice, the dynamic bound state is formed by the injected electrons, therefore the critical temperature in MATBG superlattice directly depends upon the bias voltages. In addition, if different bias voltages are used to produce the same electron density required for achieving superconductivity in MATBG superlattice. For instance, half of the electron density is created by using a 200 meV bias voltage; the other half of the electron density is injected by using a 300 meV bias voltage. In this case, the superconductivity cannot be achieved because there is no unified dynamic bound state in MATBG superlattice, but there is the "Mott insulator". Based on the definition about the type of superconductors given in the Part 1[1], the superconductivity occurred in MATBG superlattice should belong to conventional superconductors, because its critical transition temperature will vary with the bias voltages and twisted angles, through which the volume of AA stacking regions will be changed.

So far, the mainstream research on superconductivity still believe that superconductivity is caused by cooper pairs, or an ordered condensed state consisted of cooper pairs, and the insulating characteristic of superconducting state results from the strong correlation among free electrons (Mott insulator). However, experimental results show that the coherence length in the MATBG superlattice is 50 nm order of magnitude, and the optimal doping density is $n_e = 1.5 \times 10^{11}$ $cm^{-2}$ [15]. The coherence length we predicted above is 45.7nm with a doping density $1.15 \times 10^{11}$ $cm^{-2}$. In both cases, there are thousands of carbon atoms between two bound electrons (or conducting electrons). No one can believe that there is indeed a force in physics, which can bind such two electrons into a pair. What is even stranger is how the interaction between such electrons can transfer the MATBG superlattice from a conductor to an insulator.

Recently, experimental results on tuning coulomb screening show that the effect of coulomb screening on electron-phonon coupling is the dominate mechanism for superconductivity in MATBG superlattices [18]. The experimental results reported by G. Trambly de Laissardiere suggest that strain and disorder play a more important role in the observation of correlated states [19].

Until now, people still keep asking whether cooper pair or ordered condensed state can lead to superconductivity, and no one can give a convincible answer. However, it is the common belief that it works in superfluid helium. In the next section, we will demonstrate that ordered condensed state couldn't explain the peculiar properties of superfluid helium.

## III  On Superfluidity

Since the discovery of superfluid helium by P. Kapitza and J. F. Allen in 1938, the peculiar properties of superfluid helium have been extensively studied in both experimental and theoretical aspects [20, 21]. F. London first proposed that the phase transition phenomena of liquid helium at $T_\lambda$ -point might be regard as due to the characteristic of the idea Bose-Einstein particles condensed to their quantum ground state [22]. London also suggested that there might be a strong connection between superfluidity and superconductivity, and both superfluidity and superconductivity were quantum mechanism representing on a macroscopic scale. However, London's proposal is based on the assumption that liquid helium is regarded as non-interacting boson gas. In fact, there exists strong interactions between atoms in superfluid helium. Thus, London's model itself has an intrinsic defect.

In 1941, Landau proposed that the properties of superfluid helium can be understood in terms of the excited states of liquid helium, so-called as phonons and rotons [23]. According to this model, the superfluid helium can be regarded as consisting of two interpenetrating component: a normal component and a superfluid component. Normal fluid has convention viscosity, carrying all thermal energy and entropy of the system. Superfluid component can flow without friction and carry no thermal energy. At zero temperature, there is only the superfluid, and at the transition temperature, $T_\lambda$ there is only the normal fluid. These two components can move independently. Since the two-fluid model failed to point out what mechanism makes the identical helium atoms behave so differently in normal and superfluid flows. Thus, it is not surprising that the two-fluid model cannot provide essential explanation for all the peculiar properties of superfluid helium.

On the other hand, the Bose-Einstein condensation (BEC) model is widely accepted in both superfluid and superconductivity fields. It is strongly believed that the phenomena of superfluidility of Boson particles and superconductivity of cooper pairs are closely related to the BEC model, in which individual particles overlap until they behave like a big entity. Then the particles acting in unison do not behave like individual one. They march in unison without colliding with each other. Thus, the coherent matter waves can persist in superfluid helium and superconductivity.

After London's proposal, in order to determine whether the BEC is indeed present in superfluid helium, the research in both theoretical and experimental fields on superfluid helium has never stopped. The experimental method is to use neutron scattering to search for the helium atoms with zero momentum. In 1995, the experimental results reported by H. R. Glyde show that even at $0^0$ K, only 7% of helium atoms can be condensed into the Bose-Einstein ground states [24]. If there is only 7% BEC, why all of the liquid helium still behave as superfluid at $0^0$K. This conflict reflects that BEC model is not the right key to solve the mysterious superfluidity and superconductivity. Supposing that 100% of the helium atoms condense into zero momentum ground state, the sharp conflict that still must be faced is that the persistent matter current in both superfluid helium and superconductors should be definitely related to the properties of the excited states, not the ground states. Therefore, we want to say that more than 80 years later, there is still not unified theory that can consistently explain the peculiar properties observed on superfluid helium. Because neither the BEC nor the two- fluid models can reveal the dynamic nature of superfluid helium caused by high-energy particles, these high-energy particles play the driving force role in both superfluid and superconductivity.

## 3.1 A new physical model of superfluid helium

Superfluid helium and superconductivity do have some features in common, which indicate that they should have similar physical mechanism. For instance, they both have an

order- disorder transition in which their heat capacities undergo discontinuous changes. In addition, the ability of superfluid helium to maintain circulating mass current in a ring-shaped container is closely similar to that of persistent electric current in superconducting ring. Based on the physical mechanism elucidated in the superconductivity section above, the coherent matter currents in both cases are generated by the dynamic process caused by high-energy particles.

We propose that superfluid helium is formed by normal liquid helium mixed with high-energy helium atoms produced during the phase transition. The core nature of this model is the function of high-energy particles, which lie in the heart of both superfluidity and superconductivity.

The well-known feature of liquid helium is that the density of liquid helium increases linearly with cooling it from 4.2 K (boiling point) to 2.176 K ($T_\lambda$). The density reaches an abrupt maximum (0.146 gm/cm$^3$) at $T_\lambda$, and then decreases slightly with cooling it from $T_\lambda$ to absolute 0 K, as shown in Figure 4. The key of solving all peculiar properties observed in liquid helium is just hidden in the density- temperature correlation characteristics of liquid helium. In the temperature zone from 4.2K to $T_\lambda$, as liquid helium is cooled, the liquid helium can reduce its energy by emitting the thermal vibration energy of the helium atoms. It follows that the volume of liquid helium must shrink, which leads to an increase in its density. The density of liquid helium reaches its maximum at $T_\lambda$ point, which means that the liquid helium has already dissipated all its thermal energy produced by atomic vibrations. In this case, if the liquid helium is forced to continue to lower its temperature, the only way it can do is to evaporate its atoms into the space above the surface of the liquid helium.

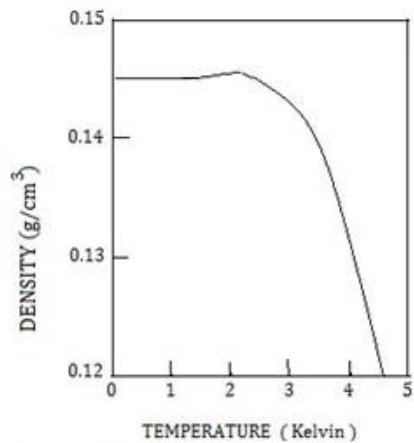

Fig. 4. The density of liquid helium as a function of temperature. The density first increasing with fall of temperature from 4.2 K to $T_\lambda$ reaches a maximum at $T_\lambda$, then slightly decreases with decreasing temperature.

If the liquid helium evaporates one atom into the space, it means that the liquid helium releases $\frac{3}{2}kT$ thermal energy from the liquid state. Here, T, is temperature of the liquid helium, k is Boltzmann's constant. In a closed container, after a violently boiling, a large number of helium atoms have evaporated into the space above liquid surface. Then helium atoms become free atoms with a kinetic energy equal to $\frac{3}{2}kT$, so the helium atoms have a thermal velocity,

$$v = (3kT/m)^{1/2} \tag{2}$$

At the transition point $T_\lambda = 2.167K$, the thermal velocity of helium atoms is approximately $1.17 \times 10^2$ m/s, which corresponds to a kinetic energy 0.28 meV. When these atoms hit the walls of container, they exert a pressure on the wall; some vaporized atoms will hit the surface of liquid helium, and return to the liquid helium as high-energy helium atoms move in the liquid helium. When the density of the gaseous helium atoms in the space above the liquid is equal to the density of the high-energy helium atoms in the liquid helium, then the boiling stops, and the dynamic equilibrium state is established. In this situation, the pressure suffered by walls of container and liquid surface is called the saturated vapour pressure. What need to point out here is that the boiling process of liquid helium at $T_\lambda$ point is completely different from the boiling process of water, the purpose of liquid helium boiling is to release its thermal energy, and the bubbles can be randomly formed at any place in the liquid helium without having any hot defects. In the meanwhile, the free surface of liquid helium is the most energetically favorable place for evaporating helium atoms into the space above liquid helium.

Since a large number of high-energy helium atoms return to the liquid helium, the free energy of the liquid helium will greatly increase. Under the saturated vapour pressure, the pressure and temperature of the entire system can be hold as constant. Thus, the incremental free energy caused by high-energy atoms can only be used to increase the volume of liquid helium. Physically speaking, the liquid helium must apply the part of its incremental free energy to increase its volume, and use the remaining part to maintain the new equilibrium state. As a result, the volume of liquid helium at below $T_\lambda$ point becomes expanded, thereby reducing the density of liquid helium, or in other words, as the temperature of the liquid helium decreases, the interatomic distance of liquid helium must be increase.

On the other hand, with lowering the temperature of liquid helium, high-energy helium atoms, through internal friction, carry normal helium atoms to form vortices, and a batch of vortices interact each other to form turbulences. A large number of vortices and turbulences naturally make the superfluid phase disorder. Now we can draw the conclusion that the order-disorder transition in liquid helium, like in superconductors, is caused by high-energy particles existing in the corresponding materials.

In the temperature zone from the $T_\lambda$ point to absolute zero, the methods used to cool liquid helium have to repeat the above process again, but the helium atoms evaporated at each process should always be the normal helium atoms, not high-energy atoms, because most of them have already become bound vortex states. After a couple of cooling processes, the free energy of liquid helium must continue to increase. As a result, the interatomic distance of the liquid helium also increased slightly. W. Dinowski et al found that at below $T_\lambda$ (1.83 $^0K$) the interatomic distance is about 10% larger than the average [25]. That is why even at absolute zero temperature liquid helium still cannot be solidified.

The increase of interatomic distance is in favor of decreasing the internal friction between helium atoms. On the other hand, the density of high-energy helium atoms continues to increase with cooling liquid helium from $T_\lambda$ to absolute zero. This effect would effectively increase the rate of shear velocity, which is inversely proportional to viscosity. Therefore, as temperature is lower than $T_\lambda$, the both effects above lead the viscosity of liquid helium to sharply decrease, but it will never become zero even at absolute zero temperature.

Based on discussions above, we can naturally draw the conclusion that superfluide helium is consisted of normal liquid helium mixed with high energy helium atoms. Keeping with this model in mind, all the peculiar features discovered in superfluid helium can be consistently explained.

## 3.2 Superfluid Helium Can Climb up the Container Wall

The strangest feature of superfluid helium is that it can climb up the walls of a container. The fact is that superfluid helium can climb over the inner walls and drop from the bottom of outside of the container until all the superfluid in the container has been siphoned out. It appears that superfluid helium can defy gravity. If you don't believe it can defy gravity - the most basic foundation of physics, then the key to solving this strange problem is hidden in Archimedes's principle, discovered about 2250 years ago, which may be the earliest principle in physics.

Before helium liquid is cooled to the superfluid transition point $T_\lambda$, the helium vapour has been already condensed on the container wall, forming a Rollin film of about 30 nm [26], which continuously covers both inner and outer walls of container. The surface tension caused by the cohesive force between the helium atoms will draw helium atoms closer together, and tends to minimize the surfaces area. So that, the energy state of helium atoms in Rollin film is more stable than that of helium atoms in container. At $T_\lambda$ point, the boiling bubbles occur randomly at anywhere of the liquid helium, but not in the Rollin films. Therefore, the liquid helium density in the Rollin films should maintain the maximum liquid helium density at the $T_\lambda$ point, while the liquid helium density in the container undergoes a slight decrease caused by high-energy helium atoms. Just this slight decrease in the density achieves the superfluid helium to climb up the walls of the container.

Let us assume that $\rho_r$ is the density of liquid helium in the Rollin film, $\rho_s$ is the density of liquid helium in the container, according to Archimedes's principle, the upward velocity of a tiny ball of liquid helium in the Rollin film is

$$V = (\rho_r - \rho_s)/\rho_s \cdot gt = at \qquad (3)$$

Where g is gravitation constant, t is time, a is acceleration. Since the density of liquid helium at the $T_\lambda$ point is not fixed, varing from 0.147 mg/cm$^3$ to 0.1462 mg/cm$^3$. Based on the data given by [27], at T=1.8K the density of the liquid helium in container is 0.145mg/cm$^3$, then the acceleration of the tiny liquid helium ball in the Rollin film is changed from 8.2 to 10.3 cm/s$^2$. In this case, it takes about two seconds for the tiny ball in the Rollin film to reach the accepted film flow velocity 20 cm/s.

Through the internal friction, the high-energy helium atom can carry the normal helium atoms with it to form a vortex, and the helium atoms in the vortex can also move with it. That is, vortices have ability to transport mass and energy with them. Thus, we can imagine that a vortex like a tiny ball carries a number of helium atoms with it to climb up the container's wall through Rollin films. When the first vortex ball arrives at the top edge of the wall of the container, it must stay there, waiting for more vortices coming to push it to over the top edge of container, then enters the Rollin film on the outside of container. From then on, the vortex in the Rollin film on the outside of the container becomes the resistance force, which will push the vortices in the Rollin film inside the container back. Therefore, the velocity of the film flow will begin to decrease linearly with more vortices moving into the outside Rollin film. Supposes that the surface level of liquid helium inside the container is higher than the outside container, and if the flow of vortices in the Rollin film outside the container moves down to the point that has the same height with the liquid helium inside container, then the velocity of the vortex flow caused by buoyant force becomes zero. Below that point, the velocity of the vortex flow in Rollin film will be controlled by the height of liquid helium in the container, that is to say, controlled by the hydrostatic process. The behavior of the vortex flow discussed above is in good agreement with the diagram of the rate of helium flowing out of container with a linearly decreased gradient measured by Danut and Mendeissohn (Figure 5, 1939) [28]. Suppose that the cylinder container with radius r is filled with superfluid helium to a height h, then the rate of liquid helium flowing out the container should be $\pi r^2$ dh/dt, here dh/dt can be regarded as the velocity of the vortex flow at the corresponding height. The slope of the straight line in the graph (Figure 5) should equal to the value of acceleration in equation (3), which only depends on the difference between the liquid helium density in the Rollin film and the liquid helium density in the container.

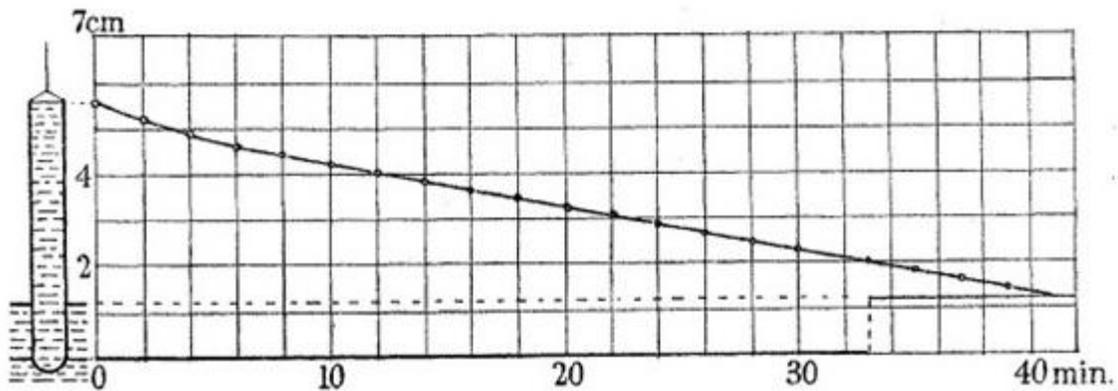

Fig. 5. Graph showing the level of helium II in an open container over time. The rate of helium flowing out of its container followed a linear gradient [28].

### 3.3 On the Fountain Effect

In 1938, Allen discovered another peculiar feature of superfluid helium, the fountain effect. Among them, a flask is connected to a reservoir of superfluid helium by a super-leak; through it, superfluid helium can pass easily, but the normal liquid helium cannot pass due to its viscose [29]. When the superfluid helium in flask is slightly heated by the heater installed in the flask, the liquid helium in flask shoots out from the opening like a fountain.

The widely accepted explanation for the fountain effect is that when the superfluid helium in the flask is heated, the superfluid helium will become normal liquid helium. In order to maintain the equilibrium fraction of superfluid helium, the superfluid helium in reservoir must rush into the flask, which will increase the pressure and cause the normal liquid helium to fountain out from the opening of flask. Obviously, this explanation is not satisfactory.

Assuming that the superfluid helium temperature in the reservoir is held at a temperature $0.0001\ ^0K$ below $T_\lambda$, while the superfluid helium in the flask is heated up to a temperature $0.0001\ ^0K$ above $T_\lambda$, a fountain effect should appear in the flask. In this case, the total energy transferred to the fountain system is about $2\ C_v\ m\ \times 10^{-4}$ Joule, here m is the mass of superfluid helium in the interface between heater and superfluid, which is close to 0.1 gram. $C_v$ is specific heat of superfluid helium at $T_\lambda$ point, which is about 10 Joules/gm.K. Thus, for starting the fountain effect, the total energy transferred into fountain system is about $2 \times 10^{-4}$ Joule. However, the energy required moving one gram of liquid helium from the surface of liquid helium in the flask to its top opening is about $0.4 \times 10^{-3}$ J. Obviously, it is impossible by using such small amount of energy to achieve the fountain effect.

According to the model, we proposed above, the superfluid helium is consisted of the normal liquid helium mixed with high-energy helium atoms. As superfluid helium in the flask is heated, the high- energy helium atoms are immediately evaporated out of the liquid helium, becoming helium atoms with an average speed about 120 m/s rushing to the opening at top of flask. In order to maintain equilibrium of the density of high-energy helium atoms between the flask and the reservoir, the high-energy helium atoms in the reservoir must continuously rush to the flask to keep the fountain over time. Thus, we can see that the energy required to fountain liquid helium out from the top opening of the flask is inherently stored in the superfluid helium itself.

### 3.4 On the Kapitza Conductance

In 1941 Kapitza, one of the founders of superfluid helium, found that when heat is transferred from a copper block to superfluid helium, there exists a temperature discontinuity at the interface between the copper block and the superfluid helium, which is often referred to as the Kapitza conductance as shown in Figure 6 [30].

There are two theories trying to explain the effect of Kapitza conductance. One is the theory of phonon radiation; the other is the theory of acoustic mismatch. The first one overestimates the value of Kapitza conductance, while the second one underestimates its value. This is because both theories fail to grasp the nature of the Kapitza conductance.

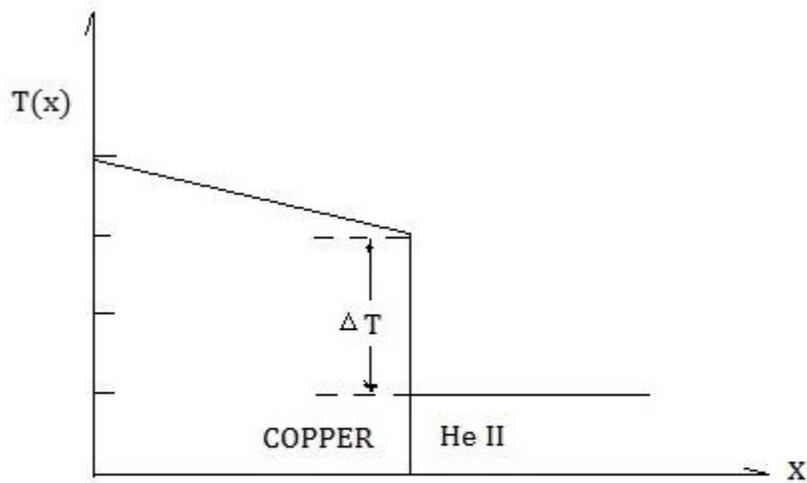

Fig. 6. Sketch of the temperature discontinuity at the interface between a solid copper block and the superfluid helium, when heater is transfered from a copper block to the superfluid helium.

The experiment results show that the Kapitza conductance only occurs when the temperature of liquid helium is lower than $T_\lambda$, for normal liquid helium ($T > T_\lambda$) the effect of Kapitza conductance is negligible.

The only difference between superfluid helium and normal helium is that there are a large number of high-energy helium atoms in superfluid helium instead of normal liquid helium. The high-energy helium atoms through the internal friction can carry the normal atoms to form the vortex bound state. In this situation, when a warm copper block is in contact with the surface of superfluid helium, there will not be enough free helium atoms at the interface, through which the heat can be transferred into the superfluid helium by atomic thermal vibrations or convections. In this case, the heat at the interface cannot be directly transferred to helium atoms bound in vortices, but it can be slowly transferred to the entire vortex as its internal energy. Thus, we can see that Kapitza conductance reflects the inherent difficulty for transferring heat from solids to superfluid helium. That is, Kapitza conductance arises from the intrinsic nature of superfluid helium, which cannot be completely eliminated through any cooling methods.

### 3.5 On Discontinuity of Specific Heat Capacity at Phase Transition

Order – disorder phase transition is always accompanied by a sharp discontinuity in the specific heat capacity around the phase transition temperature $T_\lambda$ (below and above) (Figure 7).

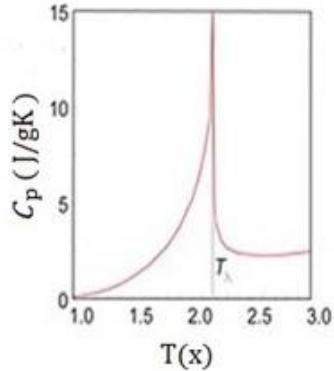

Fig.7. The specific heat capacity of liquid helium as a function of temperature. As temperature drops past the $T_\lambda$ point, a discontinuity occurs in the specific heat capacity, which also accompanied with an order-disorder phase transition in liquid helium.

This critical phenomenon has been a long-standing problem in superfluid and superconductors. Despite of thousand of articles already devoted to it, there is still no reliable answer to this century- long challenge [31].

Based on the model mentioned in the superconductivity section above, the answer to this long-standing problem is straightforward. For instance, when a superconducting material transitions from a normal state to a superconducting state, all high- energy electrons trap themselves into potential wells through dynamic interaction, forming dynamic bound states. It can be seen that the material undergoes an order – disorder transition. Now, if the specific heat capacity of the material is measured from a temperature below the transition temperature to the transition point, then at the transition point, the heat capacity consists of two parts. One is the specific heat capacity contributed by the normal state of the material, the other is the extra energy required to excite all bound electrons in potential wells to their free-electron states, and the extra energy that makes the specific heat capacity discontinuous at the phase transition point of the superconductor. Conversely, if the specific heat capacity is measured from a temperature above the transition temperature to the transition point, the specific heat capacity only results from the normal state of the material.

The mechanism responsible for the asymmetry of the specific heat capacity of liquid helium near the $T_\lambda$ point (below and above) is quite similar to that occurs in superconductors. Since the specific heat capacity of superfluid helium at the transition point, $T_\lambda$ results from two contributions. One is from normal liquid helium, the other is the energy used to excite all vortices and turbulences bound states into a state in which high energy helium atoms move freely in normal liquid helium. It is the second energy that causes the sharp discontinuity in the specific heat capacity experienced near the phase transition point.

What we need to point out that, in principle, the energy used to excite the bound electrons (or atoms) into free electrons (or atoms) only has a slight effect on the thermal energy or atomic vibration energy that determines the temperature of the material. When high-energy helium atoms stay in their free-movement state, the heat in superfluid helium can be transferred mainly through convective motion of high-energy helium atoms, which is driven by density gradient rather than temperature gradient. That is why superfluid helium has an extremely high thermal conductivity.

### 3.6 On Persistent Coherent Mass Flow in Ring-Shaped Container

Now, we begin to deal with the most challenging problem, the ability of superfluid helium to sustain persistently a coherent mass flow in a ring-shaped container. This characteristic is very similar to the persistent electric current in a superconducting ring. However, the mechanism of generating a persistent mass flow of superfluid helium is much more complicated than that of causing superconductivity in solid materials.

Now let us consider a cylinder that keeps liquid helium above the transition temperature, and then sets it to make stationary rotation at a constant angular velocity $\Omega$, the entire liquid helium rotates with the cylinder like a solid body. As the liquid helium is cooled down to below $T_\lambda$, a large number of high-energy helium atoms will be created in the superfluid helium. In the rotating cylinder, these high energy helium atoms will become vortices with a linear velocity, $\vec{v}(r) = \vec{\Omega} \times \vec{r}$, where r is the distance from a helium atom to the rotation axis of the cylinder. Thus, the rotation energy of the vortex is proportional to the distance from its position to the axis of rotation of the cylinder. Thus, all of vortices have the same vorticity, $\nabla \times \vec{r} = 2\Omega$. Since all vortices have the same rotation direction as that of the cylinder, based on the Bernoulli principle, the interactions among vortices must repel each other. Consequently, most of vortices will spontaneously arrange themselves to form a lattice structure in favor of having the lowest energy under a stationary rotation. According to experimental results, the lattice structures formed by vortices are likely to fall into the simple cubic or hexagonal lattices. In fact, in a rotating cylinder, it is impossible to arrange vortices to form standard lattice structures as in crystalline. A possible way to solve this problem is to assume that the lattice structure created by vortices is the distorted cubic lattice as shown in Figure 8a. For the sake of convenience, assuming that the lattice structure formed by vortices is a distorted cubic lattice, and then the remaining vortices will occupy the centers of the distorted cubic cells through the dynamic interaction process. That is to say, the vortex uses its kinetic energy to push its nearest adjacent eight vortices moving outward to trap itself in the center of a deformed cubic cell as shown in Figure 8. The entire trapping process should keep the total energy being a minimum. Supposing there is a circular chain composed of distorted cubic cells with a radius r in a plane perpendicular to the rotation axis of the cylinder, and then the length of the circular chain ($2\pi r$) must equal to an integer multiple of coherence length. The coherence length must be, an even number times lattice constant. For instance, if the length of the circular chain $2\pi r = 2na$, here n is an integer, a is lattice constant, 2a is the coherence length, which means that there is a vortex bound state for every two distorted cubic cells in the circular chain as shown in Figure 8a and 8b.

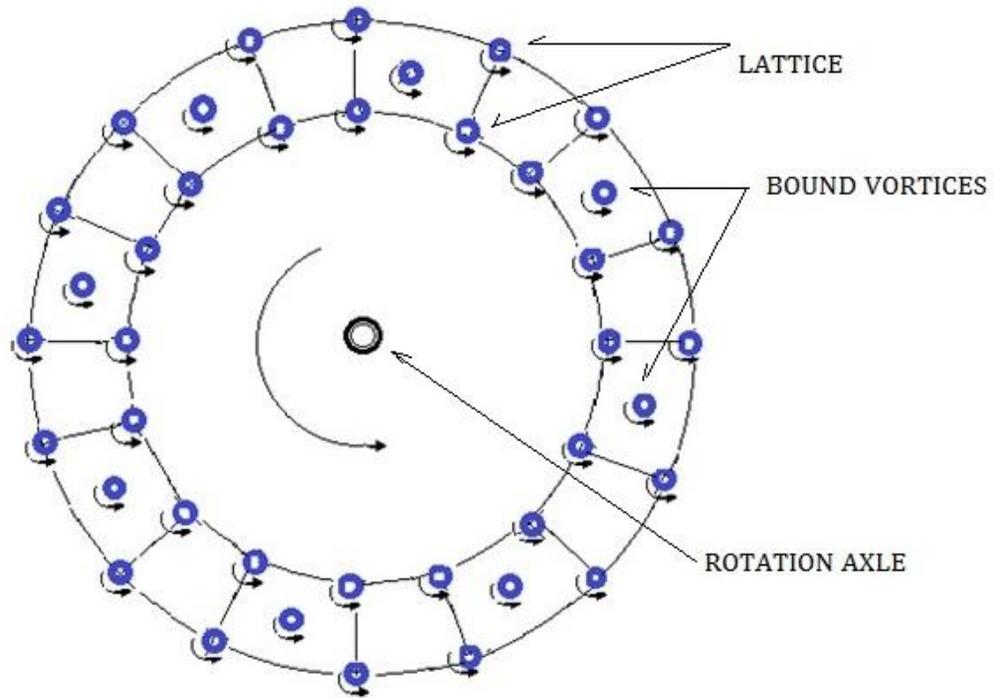

Fig. 8a. The top view of a circular chain made up of distorted cubic unit cells.

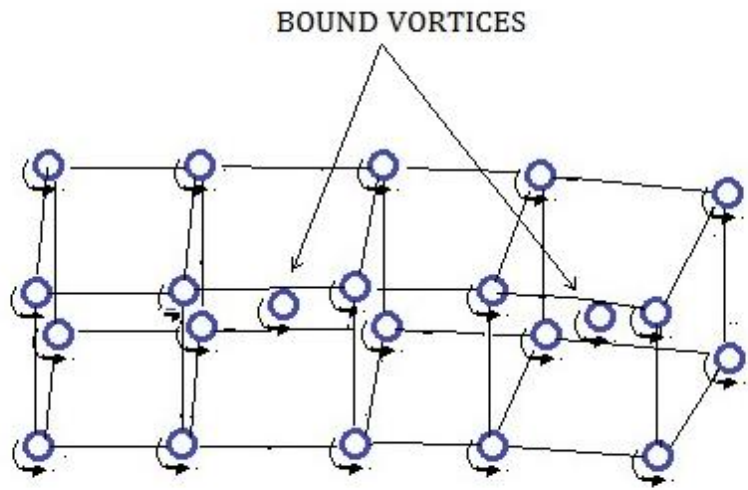

Fig. 8b. The side view of the partial circular chain in fig. 8a.

We need to emphasis that all vortex bound states in the circular chain must be created by vortices with the same energy. In the other wards, these vortices must have the same distance from their positions to the rotation axis of cylinder, since their energy is proportional to $(r\Omega)^2$.

Now, when the stationary rotation cylinder is brought to a stop, similar to a rigid body, both the liquid helium and the lattice formed by vortices in the cylinder will stop rotating synchronously. On the other hand, when the cylinder was stopped suddenly, all the trapped vortices states experience an impulse momentum, which will turn each bound vortex into a free one moving along the circular chain. At the same time, the circular chain begins to relax towards its equilibrium state. When the each vortex reaches the center of its nearest distorted cubic cell and traps itself there, then both vortex wave and lattice distortion wave have traveled a half wavelength along rotation direction. Since each vortex obtain an impulse momentum from the cylinder stop, each vortex will exert an additional impulse momentum on its around lattice in the new bound state. So that, the lattice distortion along the rotation direction will also undergoes a corresponding increase. Since the energy of trapped vortices cannot maintain the additional distortion, as a result, the vortex lattice must restore towards its equilibrium state. At the same time, every vortex withdraws its energy and impulse momentum, moving forward to the next nearest distored cubic cell. After all vortices trap themselves into the corresponding distorted cubic cells, both the circulating mass flow and lattice distortion wave have traveled one wave length along the circular chain. From then on, the circular chain will continue to repeat the above dynamic process. The mass current will coherently move with lattice distortion waves without being dissipated.

From the discussions above it follows that if the total number of distorted cubic cells contained in a circular chain is an odd number, the chain will lose the ability to maintain a persistent mass flow. It further means that half of the superfluid helium in the rotation container has no opportunity to support the persistent mass flow. In addition, even if only one vortex bound state in the circular chain is replaced by a vortex with a different energy from the original vortex, the mass flow in the circular chain must be decayed. In short, there are a number of factors, which can destroy the persistent mass flow. However, there is no doubt that the persistent mass flows do exist in superfluid liquid helium, but the conditions for realizing the persistent mass are very critical.

## IV  Conclusions

The experiment results obtaining from various materials and different conditions should be a touchstone for judging the correctness of theories. Based on discussions on superconducting properties discovered in various materials, we are forced to conclude that the physical mechanism of superconductivity proposed in Part I [1] is the only way to handle superconducting properties in all superconductors. The mechanism leading to metal-insulator transition in metal oxides or "Mott insulator" is just a part of the mechanism of superconductivity.

So far, the mainstream research on superconductivity still believe that superconductivity is caused by Cooper pair or the ordered condensed state made of Cooper pairs, and the insulating characteristic of superconducting state in all superconductors is attributed to the strong correlation between free electrons, or "Mott insulator". However, the experimental result of MATBG superlattice shows that the coherence length is 50 nm order of magnitude at the optimal doping density $n_e = 1.5 \times 10^{11}$ cm$^{-2}$. Our predicted coherence length is 45.7nm, corresponding to a doping density of $1.15 \times 10^{11}$ cm$^{-2}$. In both cases, there are thousands of carbon atoms between the two bound electrons. No one would believe that there is indeed a force in physics, which can bind such two electrons into a pair. Even stranger is how the interaction between such electrons transfers the MATBG superlattice from conductor into insulator.

Based on superconducting process discussed previously, the superconducting state itself is an insulating state, which is caused by the dynamic bound state of high-energy electrons trapping into three-dimensional potential wells. Therefore, the superconductivity in all of superconductors must start from an insulating state, which is the nature philosophy.

The physical concepts, such as Cooper pair and Mott insulator, they are all man-made and unable to unify with fundamental laws of physics. More than a half century has been wasted on this kind of wrong concepts. Until now, people still keep asking whether Cooper pairs or their ordered condensed state can caused superconductivity. We believe that no one can give a convincible answer, because the reality described by such concepts does not exist in nature.

We propose that superfluid helium consists of normal liquid helium mixed with high-energy helium atoms. Based on this new model, all peculiar features observed in superfluid helium have been consistently interpreted in the text. The core nature of this model is the high-energy helium atoms, which is generated in a phase transition at the $T_\lambda$ point. The high-energy particles in both superconductors and superfluid helium play driving force role and thereby dominate their properties.

What we need to learn from the development history of superconductivity and superfluid helium is that choosing the right physical models is critical for all research efforts. If the physical model chosen to describe a particular phenomenon cannot be unified with the laws of physics, such as the Cooper pair model for superconductivity and BEC macroscople quantum state for superfluid helium, then it is impossible to truly understand the nature of a specific phenomenon.

**Acknowledgement**

We would like to thank the authors of the articles cited in this article. The senior authors of this paper are particularly grateful to Professors Heribert Wiedemeier, John Schroeder, and Peter Persans for their support during their work at Rensselaer Polytechnic Institute in Troy, NY. U.S.